\documentclass[prb,twocolumn,showpacs,superscriptaddress,preprintnumbers,amssymb,hyperlink]{revtex4-1}
\usepackage{tabularx,xcolor}
\usepackage{graphicx,adjustbox} % Required for inserting images
\usepackage{amsmath}
\newcommand{\cL}{ {\cal L} }

\newcommand{\sgn}{{\rm sgn}}

\newcommand{\ra}{\rightarrow}

\newcommand{\ii}{\mathrm{i}}

\newcommand{\U}{\mathrm{U}}

\newcommand{\beq}{\begin{equation}}
\newcommand{\eeq}{\end{equation}}
\newcommand{\beqn}{\begin{eqnarray}}
\newcommand{\eeqn}{\end{eqnarray}}

\begin{document}

%\title{Fractional Quantum Anomalous Hall Crystal and its Proximate Phases}
\title{Intertwined fractional quantum anomalous Hall states and charge density waves }

\author{Xue-Yang Song}
\affiliation{Department of Physics, Massachusetts Institute of Technology, Cambridge, MA 02139}

\author{Chao-Ming Jian}
\affiliation{Department of Physics, Cornell University, Ithaca, New York 14853}

\author{Liang Fu}
\affiliation{Department of Physics, Massachusetts Institute of Technology, Cambridge, MA 02139}

\author{Cenke Xu}
\affiliation{Department of Physics, University of California, Santa Barbara, CA 93106}

\begin{abstract}

Motivated by the recent experimental breakthrough on the observation of the fractional quantum anomalous Hall (FQAH) effects in semiconductor and graphene moir\'{e} materials, we explore the rich physics associated with the coexistence of FQAH effect and the charge density wave (CDW) order that spontaneously breaks the translation symmetry. We refer to a state with both properties as ``FQAH-crystal''. 
%\cx{around a phase with coexisting FQAH effect and CDW order, which is a state we refer to as ``FQAH-crystal"}. 
We show that the interplay between FQAH effect and CDW can lead to a rich phase diagram including multiple topological phases and topological quantum phase transitions at the same moir\'e filling. In particular, we demonstrate the possibility of direct quantum phase transitions from a FQAH-crystal with Hall conductivity $\sigma_H = - 2/3$ to a trivial CDW insulator with $\sigma_H = 0$, and more interestingly, to a QAH-crystal with $\sigma_H= -1$.    

\end{abstract}

\date{\today}

\maketitle

\section{Introduction}

%{\bf Motivation.....} 

The recent advance in the fabrication and control of two-dimensional (2D) van der Waals heterostructures has enabled the development of moir\'{e} superlattices that feature tunable mini-bands. %The moir\'{e} systems formed from stacking and twisting graphene, monolayer cuprates and transition metal dichalcogenides (TMD), have led to exciting discoveries of strongly correlated quantum phenomena, such as unconventional superconductivity and correlated insulators\cite{cao2018,kim2022,li2021continuous}. %\cx{(we need a reference for twisted monolayer cuprates, if we do want to mention it)}. 
Topologically nontrivial mini-bands in moir\'e materials provide an ideal avenue to search for topological states of matter. As an example, evidence of fractional quantum Hall effect in fractionally filled moir\'e band was observed in twisted bilayer graphene under a  magnetic field above $\sim 5$T or higher\cite{xie2021fractional,spanton2018observation}. More recently, thermodynamic and transport measurements revealed the fractional quantum anomalous Hall (FQAH) effect at zero magnetic field in twisted TMD homobilayers\cite{cai2023signatures,zeng2023integer,park2023observation, xu2023observation} and rhombohedral pentalayer graphene/hBN superlattice\cite{lu2023fractional}.  

%At zero magnetic field, the strong atomic spin-orbit coupling in the TMD system locks the valley and spin quantum numbers. Spontaneous time reversal breaking and ferromagnetism result from valley polarization, due to small moir\'{e} bandwidth and nontrivial band topology\cite{wu2019topological,devakul2021magic,crepel2023fci,wang2023fractional}, across a range of hole filling $|n|=1.2\sim 0.4$. Signatures of the Hall plateaus with Hall conductivity quantized to integer (in units of $e^2/h$) and a fraction have both been identified in optical, thermodynamic and transport measurements\cite{cai2023signatures,wang2023fractional,zeng2023integer}. 

The discovery of FQAH effect points towards a fertile ground for studying strong interaction effect in topological moir\'e bands of 2D materials. %, and paves the way for even more exotic topological orders such as the non-Abelian states. 
Twisted TMD homobilayers feature spin-valley-locked moir\'e bands with opposite Chern numbers in the two valleys \cite{wu2019topological, yu2020giant, devakul2021magic}. At finite carrier density, Coulomb interaction drives spontaneous valley polarization, and FQAH effect is anticipated at fractional fillings of the valley-polarized Chern band  \cite{devakul2021magic, li2021spontaneous, crepel2023fci}. Interaction induced FQAH states in Chern bands are also known as (zero-field) fractional Chern insulators in the literature \cite{sun2011nearly,tang2011high, neupert2011fractional,sheng2011fractional, regnault2011fractional}.   
The highly tunable nature of moir\'{e} systems, with abundant tuning knobs such as twist angle, displacement field, electrostatic doping and gate screening, offers a large parameter space to explore FQAH states and proximate phases\cite{duran2023pressure,  wang2023fractional,reddy2023fractional, qiu2023interaction, crepel2023chiral, goldman2023zero, dong2023composite, 
liu2023gate, xu2023maximally,reddy2023toward, song2023phase,   yu2023fractional, abouelkomsan2023band}. %Theory predicts a plethora of correlated or symmetry breaking phases (e.g. charge density wave), and continuous phase transitions pertinent to the QAH phases in TMD homobilayers\cite{goldman2023zero,dong2023composite,song2023phase}. Lesser attention has been given to the possibility of coexistence of charge density wave(CDW) and fractional QAH (FQAH) effects.

In this work, we explore the rich physics of a state with coexisting FQAH and CDW (a state that we refer to as FQAH-crystal), and its proximate phases that occur at the same moir\'e band filling under zero magnetic field.
%we explore the interplay between FQAH states and charge density waves (CDW) that occur at the same moir\'e band filling under zero magnetic field. 
Our study is motivated by the observed phase transition at hole filling $\nu=-2/3$ in twisted bilayer MoTe$_2$  under a displacement field from an FQAH state with quantized Hall conductance $\sigma_H=-2/3$ (in the unit of $e^2/h$) to an insulating state~\cite{park2023observation,xu2023observation}. % that \cx{could be a trivial insulator} with CDW.  
We consider a scenario in which the FQAH state spontaneously breaks the translational symmetry of the moir\'e lattice. We call this state a FQAH-crystal, analogous to the notion of ``Hall crystal" introduced in Ref.~\onlinecite{hallcrystal}. Our consideration of FQAH-crystal is partly  
motivated by recent numerical finding of a softened magneto-roton gap in $\sigma_H=-2/3$ FQAH states~\cite{reddy2023fractional}, which suggests incipient CDW order with tripled $\sqrt{3} \times \sqrt{3}$ moir\'e unit cells (as illustrated in Fig.\ref{fig:cdw_bz}) at experimentally relevant twist angles.     

We show by field theory analysis that a variety of strongly correlated phases can be found in the vicinity of  FQAH-crystal. These include a trivial CDW insulator with $\sigma_H=0$ and a $\sigma_H=-1$ QAH state with CDW order, which we call QAH-crystal. While the QAH-crystal phase has been proposed in moir\'e systems under the name of topological charge density wave~\cite{polshyn2022topological, pan2022topological, xu2023maximally}, its connection to FQAH physics  \cite{kol1993fractional, kourtis2014combined, dong2022many} and phase transitions have received little attention before.   
All the phases considered in this work possess the same type of CDW order, but are distinguished by their different topological properties.

%The state with Hall conductivity $|\sigma_H| = 2/3$ (in unit of $e^2/h$) at filling $\nu= - 2/3$ of a Chern band with Chern number $C = \pm 1$, i.e. on average $2/3$ holes per moir\'{e} unit cell. The state we construct would naturally have a topological order that coexists with a CDW that spontaneously breaks the translational symmetry and enlarges the unit cell. In our construction the $\sigma_H = - 2/3$ state can be naturally driven into several other phases, which include a trivial insulator phase with $\sigma_H = 0$; integer quantum Hall states with $\sigma_H = -1, + 2$; a ``neutral" topological order with $\sigma_H = 0$, etc. For all these states we present different formalisms that are proven to be equivalent to each other. More specifically, the $K-$matrices from these formalisms are equivalent to each other through similarity transformations. 
%These formalisms include the parton construction, and a composite fermion construction. 
%We also show that the $K$-matrix of these two formalisms are equivalent to each other. 

We further show that direct and (potentially) continuous phase transitions between these topologically distinct phases are theoretically allowed. 
Interestingly, these phase transitions can be described by (2+1)D quantum electrodynamics (QED) with a Chern-Simons (CS) term of the $\U(1)$ gauge field coupled to either fermionic or bosonic charges. %For example, experimentally it was observed that there is a continuous quantum phase transition tuned by the displacement field between the $|\sigma_H| = 2/3$ FQHA state and an insulating state $\sigma_H = 0$~\cite{wang2023fractional}. 
In our theory, the transition from the FQAH-crystal with $\sigma_H = - 2/3$ to the trivial CDW insulator is described by a fermionic QED with two flavors Dirac fermions at low energy coupled with a U(1) gauge field with a CS term at level-$1/2$.  On the other hand, its transition to a QAH-crystal with $\sigma_H = - 1$ is described by either bosonic or fermionic QEDs. These two descriptions are {\it dual} to each other based on the boson-fermion duality web that was actively discussed in recent years~\cite{wudual,SEIBERG2016,wang2017deconfined,song2022deconfined}. 

We note that direct transitions between a standard FQAH state (without any spontaneous symmetry breaking) and exotic CDW states, including with topological order, were studied in Ref.~\onlinecite{song2023phase}, whereas our work starts from a FQAH-crystal (with $\sigma_{H}=-2/3$ and CDW order), and obtains different proximate phases. In particular, we highlight the possibility of a QAH-crystal (with $\sigma_H=-1$ and CDW order) as a proximate phase and a direct phase transition between FQAH-crystal and QAH-crystal at $\nu=-2/3$.   
%The difference from our theory will be discussed in section III of the current manuscript.

%We describe, at filling $\nu=-2/3$, the theory for a (fractional) quantum hall state coexisting with charge-density-wave (CDW) which triples the unit cell or a CDW insulator, with hall conductivity $\sigma_H=2/3,1,0$, respectively, and construct the transition among these states. 

\section{Phases at $\nu=-2/3$}

The most prominent state observed experimentally in the homobilayer TMD moir\'{e} system is the $\sigma_H = \pm 2/3$ FQAH state at hole filling $n_h= 2/3$, i.e., at hole density of 2/3 per moir\'{e} unit cell. Since this state is shown to be fully spin/valley polarized by magnetic circular dichroism measurements, in the following we consider a spinless electron system at charge density $\nu=-2/3$. 

\begin{figure}[tb]
\begin{center}
\adjustbox{trim={.14\width} {.5\height} {.45\width} {.16\height},clip}
{\includegraphics[width=2\linewidth]{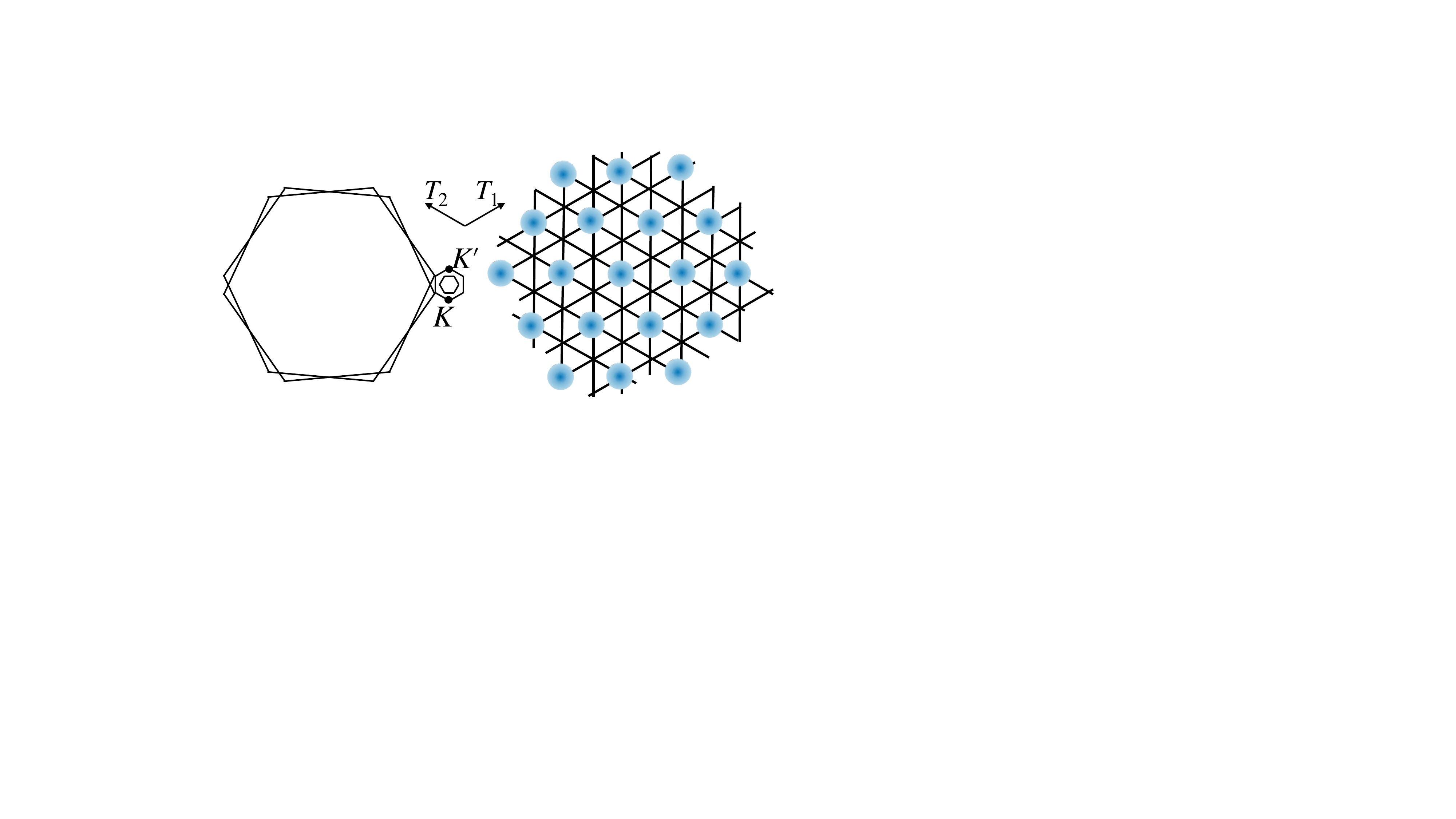}}
\end{center}
\caption{ (Left) The moir\'{e} Brillouin zone formed when twisting the original Brillouin zone (two large hexagons) of each TMD layer. When charge density wave forms that triples the unit cell, schematically shown on the right on the moir\'{e} superlattices, the Brillouin zone is further folded to the center smallest hexagon on the left.}
\label{fig:cdw_bz}
\end{figure}

Throughout the discussion below we postulate the presence of a charge density wave (CDW) which triples the unit cell (examples shown in Fig~\ref{fig:cdw_bz}), such that the holes are at integer filling with respect to the enlarged moir\'{e} unit cell. %CDW order   
%The CDW is most naturally obtained by condensing excitons at $K,K'$ momenta of the triangular Brillouin zone. Exact diagonalization results lended support to the presence of CDW as one tunes the displacement field. 
We will show that under tripling of the unit cell, it is natural to construct phases with Hall conductivity $\sigma_H=-2/3$, $-1$, $0$ in a unified formalism. %Here $C$ denotes quantized Hall conductivity in unit of $e^2/h$, or equivalently, the many-body Chern invariant of a gapped phase. 
Furthermore, there can be direct and (potentially) continuous quantum phase transitions between any of the two states mentioned above, though a direct transition between the $\sigma_H=-2/3$ state and the trivial insulator with $\sigma_H = 0$ requires certain discrete space-time symmetries.
%i.e. reflection combined with time reversal $RT$, 
%one could realize a direct, continuous transitions between any two of the above phases.

%The phases we construct can be understood from the perspective of composite fermions, . 

For the purpose of constructing these phases and describing their properties, it is convenient to use the standard parton construction. One can formally write the hole operator as $c=\Phi f$, where the bosonic parton $\Phi$ carries the physical electric charge, and the charge-neutral parton $f$ is a fermion. The electric charge can actually be assigned arbitrarily between $\Phi$ and $f$, which should not change the final physics. The parton construction formally enlarges the Hilbert space of the holes, which can be remedied by coupling $\Phi$ and $f$ both to an internal dynamical $\U(1)$ gauge field $a$, with charge $\pm 1$ respectively. The dynamical $\U(1)$ gauge field enforces a local constraint which equates the local density of $f$ to that of $\Phi$. The physical state of holes is obtained by enforcing the relation of hole density to that of the partons, i.e., $\nu_h = \nu_\Phi =\nu_f= 2/3$ with respect to the original moir\'{e} unit cell. 
Importantly, in the presence of a CDW order that triples the unit cell, both the holes and partons are  at integer fillings with respect to the enlarged unit cell.

\begin{figure}[tb]
\begin{center}
\adjustbox{trim={.14\width} {.5\height} {.45\width} {.12\height},clip}
{\includegraphics[width=2\linewidth]{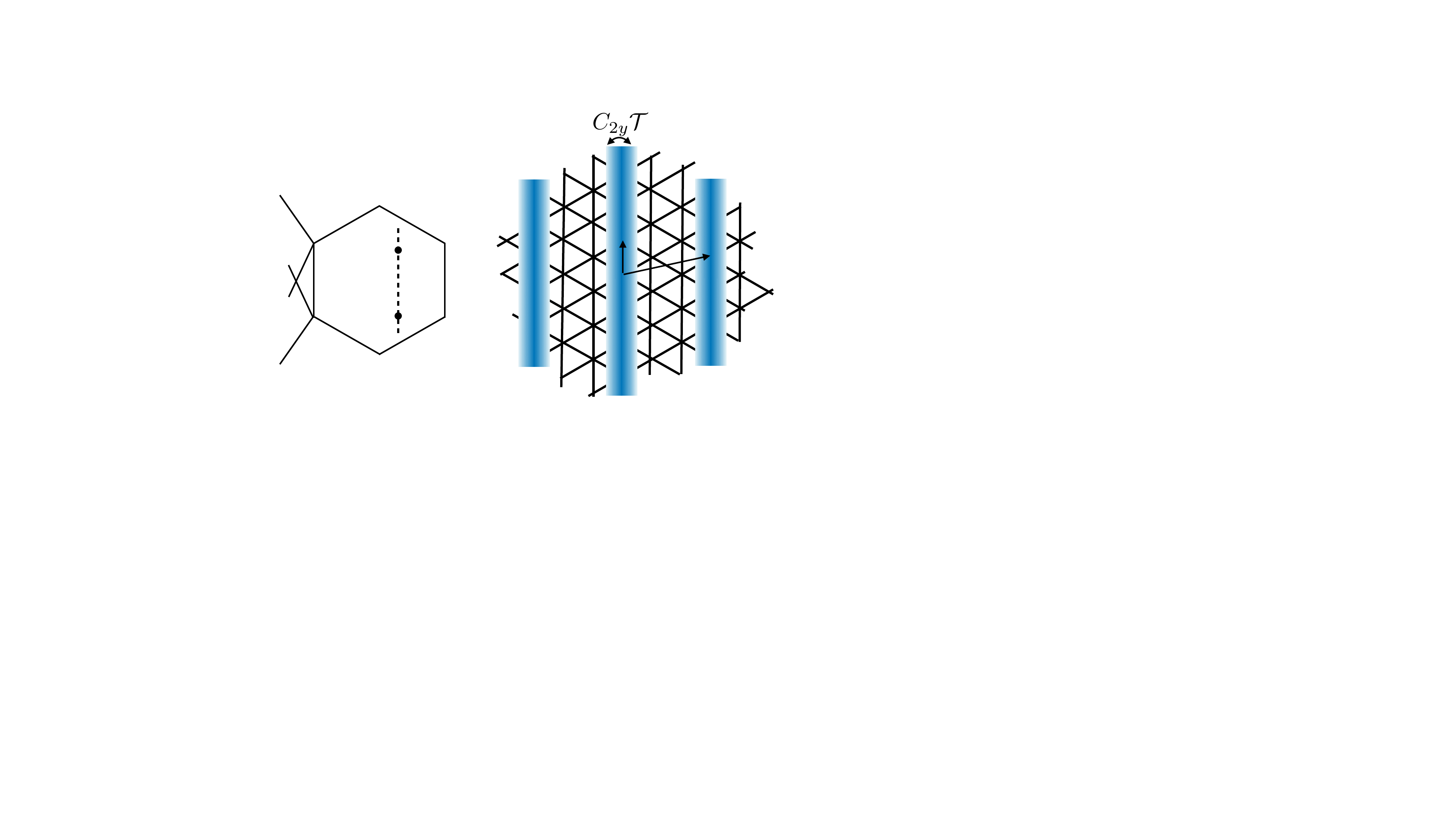}}
\end{center}
\caption{ (Left) The zoomed in moir\'{e} Brillouin zone with the symmetry $C_{2y} \mathcal{T}$ shown that protects degeneracy of $2$ generic Dirac cones(black dots) aligned along vertical axis. The arrows indicate the action of $C_{2y}$ which flips the horizontal axis. When combined with time reversal, $C_{2y} \mathcal{T}$ flips the vertical axis of the Brillouin zone and hence protects the degeneracy of Dirac cone as plotted. When charge density wave forms a stripe pattern schematically shown on the right, the symmetry is preserved and a direct transition from $\sigma_H=-2/3\rightarrow 0$ could be realized. %\cmjCom{Reminder: Change $C_{2y} T$ to $C_{2y} \mathcal{T}$ in the plot}
}
\label{fig:stripe_bz}
\end{figure}

\subsection{Phases tuned by fermionic parton $f$}

%In the following we will construct a series of states by making $\Phi$ a $\nu = -1/2$ bosonic fractional quantum Hall state. Each state can also be equivalently constructed employing the more traditional composite fermion picture through flux attachment~\cite{Jain}, though we would like to stress that the existence of the FQAH states do not rely on any external magnetic field. In the next subsection we will also discuss phases obtained by tuning the physics of bosonic parton $\Phi$. 

In the following we will construct a series of states by making $\Phi$ a bosonic fractional quantum Hall state with Hall conductivity $-1/2$. Each state can also be equivalently constructed employing the composite fermion picture through vortex %\cx{(should it be flux attachment?)} 
attachment. As is well known in the context of Landau level systems, composite fermions experience a modified residual magnetic field, and prominent fractional quantum Hall states are formed  at integer filling of composite fermion Landau levels ~\cite{Jain}. As we show later, in (moir\'e) lattice systems, the mean-field state of composite fermions allows much richer possibilities, leading to a series of new states\cite{song2023emergent}.       
%though we would like to stress that the existence of the FQAH states do not rely on any external magnetic field.

%\begin{enumerate}

{\bf --- $\sigma_H = - 2/3$ State: }

The $\sigma_H = - 2/3$ state at filling $\nu = - 2/3$ is the most prominent state observed experimentally in the homobilayer TMD moir\'{e} system. This state can be constructed naturally using the parton formalism, where $\Phi$ and $f$ each forms its own ``mean field state". With the assumption of the existence of a background charge density wave that triples the unit cell, the original Chern band in the moir\'{e} Brillouin zone (BZ) would split into three bands in the folded moir\'{e} BZ, and the fermionic parton $f$ would fill two out of the three bands due to its $2/3$ filling. One natural way to construct the $2/3$ state, is for the bosonic parton $\Phi$ to form a $\nu = -1/2$ Laughlin state, and at the mean field level the fermionic parton $f$ fills two low energy bands in the folded BZ with Chern numbers $+1, +1$. As we mentioned previously, the $\sigma_H = -2/3$ state constructed here has FQAH effect as well as spontaneous translation symmetry breaking, which we refer to as FQAH-crystal.  Later, we will demonstrate with a composite fermion construction that, the existence of the $\sigma_H = - 2/3$ state itself %does not have to break the translation symmetry; but the direct transition to other states within our formalism would enforce the breaking of translation.  
does not have to break the translation symmetry. However, all of the nearby states within our formalism must break the translation symmetry. This observation motivates us to focus on the scenario where the translation symmetry is already broken in the  $\sigma_H = - 2/3$ state.

 Here we would like to demonstrate that the parton construction given above is a natural state for holes at filling $2/3$ of the moir\'{e} unit cell. We note that the flux $\phi_\Phi$ per moir\'e unit cell felt by the parton $\Phi$ is not necessarily equal to the physical flux seen by holes $\phi_h$, due to the internal gauge field $a$ coupled to both $\Phi$ and $f$. In the continuum, the total fluxes seen by the hole and the partons should in general obey the relation $\phi_h = \phi_\Phi + \phi_f$. 
%(On a lattice this relation should still hold mod $1$). 
In twisted semiconductor bilayers\cite{wu2019topological, yu2020giant, morales2023magic} and other continuum systems\cite{paul2023giant} where holes fill a valley polarized Chern band, despite being at zero magnetic field, the holes experience an effective flux $\phi_h=-1$ per moir\'{e} unit cell produced by the periodic skyrmion spin (or layer pseudospin) texture in real space\footnote{we are not discussing a tight-binding model on the moir\'{e} scale, hence the magnetic flux cannot be removed through large gauge transformation}, and we could further set $\phi_\Phi=-4/3$ to allow $\Phi$ to form a Laughlin $\nu=-1/2$ state. This leaves $\phi_f=1/3$ and $\nu_f=2/3$. The fermionic partons hence are naturally allowed to fill two Landau levels, equivalent to filling two bands with Chern number +1. In fact, with the CDW order that triples the unit cell that we postulate, the $f$ feel $\tilde \phi_f = 1$ with a filling $\tilde\nu_f = 2$ per enlarged unit cell, hence $f$ could naturally form an insulator with total Chern number $C=2$.

In terms of the Chern-Simons theory, this state corresponds to the following Lagrangian 
\beqn
\cL &=& - \frac{2}{4\pi} b \wedge d b + \frac{1}{2\pi} b \wedge da - \frac{1}{4\pi} a \wedge (a_1 + a_2) \cr\cr &+& \sum_{i = 1,2} \frac{1}{4\pi} a_i \wedge d a_i. \label{cs1} 
\eeqn
The gauge field $b$ is the ``dual" of the current of the bosonic parton $\Phi$; $a_1$ and $a_2$ are the dual of the fermionic parton $f$ that fills the two Chern bands with Chern number $(+1, +1)$; $a$ is the gauge field that couples to both $\Phi$ and $f$. The CS Lagrangian Eq.~\ref{cs1} can also be written in a more compact form using the $K-$matrix~\cite{wenzee}: \beqn \cL = \frac{1}{4\pi} K_{2/3, IJ} a^I \wedge d a^J , \eeqn where \beqn 
K_{2/3} = \begin{pmatrix} 
-2 & 0 & 0 & 1 \\ 
0 & 1 & 0 & -1 \\
0 & 0 & 1 & -1 \\
1 & -1 & -1 & 0
\end{pmatrix}, \label{Kmatrix1}
\eeqn and $a^I = (b, a_1, a_2, a)$. We will hereafter abbreviate $a\wedge db$ as $adb$ without loss of clarity. The topological ground state degeneracy is given by the determinant of $K_{2/3}$, which in this case is $3$. To derive the Hall conductivity of this state, we need to couple $a^I$ to the external electromagnetic field $A$, through a ``charge vector"~\cite{wenzee}. In the current construction, the charge vector is $v=(1, 0, 0, 0)$, meaning that only the bosonic parton $\Phi$ carries electric charge +1. By integrating out all the dynamical gauge field $a^I$, one can show that the total Hall conductivity of the state is $\sigma_H = - 2/3$, i.e. $\sigma_H=v K^{-1} v^T=-2/3$. 

An alternative picture is that, the $2/3$ state can be viewed as holes at filling $1$, forming a $\nu = -1$ integer quantum hall state, together with electrons at filling $1/3$, forming a $\nu=1/3$ Laughlin state. The $K-$matrix for this construction is \beqn K'_{2/3}=\begin{pmatrix} -1&0\\0&3\end{pmatrix} \label{Kp}\eeqn with the charge vector $v=(1, -1)$. The first diagonal element of $K'_{2/3}$ describes the $\nu=-1$ quantum hall state and the second diagonal element describes $\nu=1/3$ Laughlin state. The $K-$matrix in Eq.~\ref{Kmatrix1} is related to the $K'-$matrix in Eq.~\ref{Kp} (up to $2$ extra fields that describe a trivial, neutral sector) by a similarity transformation in $SL(4,Z)$:
%\cmjCom{The bottom right $2\times 2$ block on the right-hand side of the first line should have extra $-1$ signs? Or we can redefine $W$}
\begin{align}
    W^T K_{2/3}W=\begin{pmatrix} -1&0&0&0\\ 0&3&0&0\\0&0&0&-1\\0&0&-1&0\end{pmatrix},
    \nonumber\\
    W=\begin{pmatrix} -1 &1&1&0\\-1&2&0&0\\0&-1&0&0\\-2&2&1&-1\end{pmatrix}.
\end{align}

The third picture of constructing the $2/3$ state is through the composite fermion (CF) and flux attachment. As we mentioned before, when the holes fill a valley polarized Chern band, the physics is topologically equivalent to integer quantum Hall state where a hole sees a $\phi_h = -1$ magnetic flux quantum through each moir\'{e} unit cell. Then a composite fermion is constructed by binding the hole with $2$-flux quanta of a gauge field $a$, i.e. the composite fermions will see total gauge flux $\phi_{cf} = \phi_h + 2 \rho_{cf}$. When the hole density is 2/3 per moir\'{e} unit cell, the density of $\phi_{cf}$ would be 1/3 flux quantum per moir\'{e} unit cell. Hence the composite fermions would naturally fill two Landau Levels of $\phi_{cf}$, and form an integer quantum Hall state with composite fermion Hall conductivity $\sigma_{cf} = 2$. When inserting an extra flux density $\delta \phi_{cf}$ into the system, composite fermion density $\delta \rho_{cf} = \sigma_{cf} \delta \phi_{cf}$ will be accumulated; and the total Hall conductivity is the ratio between the extra density of composite fermions and the density of extra magnetic flux: $\sigma_H = \delta \rho_{cf}/\delta \phi_h = \sigma_{cf}/(1 - 2 \sigma_{cf}) = - 2/3$. This composite fermion construction for the $2/3$ state does not break translation. The composite fermion picture for understanding the observed FQAH states in twisted bilayer MoTe$_2$ is strongly supported by recent numerical studies \cite{goldman2023zero, dong2023composite, reddy2023toward}.      
%However, the description of proximate states with $\sigma_H=0,-1$ within the CF framework will need to break translations as will be discussed in detail later.}

To formally implement this flux attachment\cite{barkeshli2012continuous}, we introduce a noncompact gauge field $b$ whose charge is the flux of $a$. We will demonstrate that a $\U(1)_{-2}$ CS term for $b$ attaches $2$ units of fluxes of $a$ to the composite fermion, which also carries charge under gauge field combination $a+A$. The Lagrangian of all the field mentioned above reads:
\begin{align}
    \mathcal L=-\frac{2}{4\pi}bdb+\frac{1}{2\pi} bda+\mathcal L_{CF}[\psi, a+A], \label{fluxattach1}
\end{align}
where the mutual CS term between $b,a$ implies that the flux of $a$ is charged under $b$, and the last term of Eq.~\ref{fluxattach1} is the CF Lagrangian capturing the physics that the CF ($\psi$) is coupled to $a+A$.

The equations of motion with respect to $b$ and $a$ lead to the following relations:
\beqn
    \frac{\delta \mathcal L}{\delta b_0}=0 \ \rightarrow \ \frac{da}{2\pi}=\frac{2db}{2\pi},\cr\cr
    \frac{\delta \mathcal L}{\delta a_0}=0 \ \rightarrow \ \rho_{cf}=\frac{db}{2\pi}.
\eeqn
Combining the equations we obtain the relation $2\rho_{cf}=\frac{da}{2\pi}$, which corresponds to the picture of flux attachment: each CF is bound with two flux quanta of gauge field $a$.

When the CF fermion $\psi$ fills Chern bands with total Chern number $C_{cf}$ (an integer), we need to introduce $|C_{cf}|$ copies of gauge fields $a_i$, which are dual to the current of the CFs:
\begin{align}
    \mathcal L_{CF}[\psi, a+A]=\textrm{sgn}(C_{cf})\sum_{i=1}^{C_{cf}} \left( \frac{1}{4\pi} a_id a_i+\frac{1}{2\pi} a_i d(a+A) \right), \label{fluxattach2}
\end{align}
where each self-CS term of $a_i$ describes the CF filling a complete Landau Level (or equivalently Chern band with Chern number $1$). The flux current of $a_i$, which is the dual of the CF current, couple to $a+A$.

For $C_{cf}=2$, after combining Eq.~\ref{fluxattach1} and Eq.~\ref{fluxattach2} we eventually arrive at {\it exactly the same} $K-$matrix as that in Eq.~\ref{Kmatrix1}, albeit the charge vector now is $(0,1,1,0)$, which corresponds to shifting the electric charge from the bosonic parton $\Phi$ to the fermionic parton $f$, and it still leads to $\sigma_H=-2/3$. The charge vector could be transformed to $(1,0,0,0)$, by relabeling $a\rightarrow a-A$. Then the two formalisms based on parton and CF yield exactly identical $K-$matrices and Hall conductivity.

%\cx{Xueyang, please add the CS theory and K matrix Lagrangian of the composite fermion construction of the 2/3 state; and show that it is equivalent to the parton construction}

{\bf ---  $\sigma_H = 0$ and $\sigma_H = -1$ states: }

To construct a trivial insulator phase with $\sigma_H = 0$, we can still fix the bosonic parton at a $\nu = -1/2$ Laughlin state, and let the fermionic parton $f$ fill two bands with Chern numbers $+1,-1$ respectively. The $K$ matrix of this state is similar to Eq.~\ref{Kmatrix1}, with the diagonal component $K_{33}$ changed to $-1$. This change will lead to the Hall conductivity $\sigma_H = 0$, without any topological degeneracy. 

An integer QAH state with $\sigma_H = -1$ and coexisting CDW order (referred to as the QAH-crystal state) can be constructed by removing the row and column of the $K$ matrix that involves the second band of the fermionic parton, meaning that the fermionic parton $f$ now fills bands with total Chern numbers $+1$. 
%\cmjCom{Comment: Instead of setting $K_{33}$ to 0, we should just erase the third row when the chern number is 0  }

In the composite fermion picture, when the CF forms a $\nu = +1$ quantum hall state, inserting a $+1$ $\phi_h$-flux quantum is accompanied by $-2$ units of fluxes of $a$, and accumulating $-1$ charge of CF since there is a total $-1$ unit of extra flux $\phi_{cf}$, i.e. the CFs form a $\nu = -1$ state with respect to $\phi_h$. This state eventually corresponds to the $\sigma_H = -1$ state. The $K-$matrix can be similarly deduced and is equivalent to that deduced from the parton construction.

The trivial insulator with $\sigma_H = 0$ corresponds to the CF forming a trivial insulator, which leads to trivial electromagnetic response. 
The $K-$matrices of the three states constructed in this subsection are summarized in table~\ref{tab:summary}.

%From the electron-hole perspective, the $\sigma_H=-1$ or $0$ state can be viewed as electrons at filling $1/3$ forming a trivial insulator or a $\nu = 1$ IQH state, on the background of a $\nu = -1$ state of holes. Note that the unit cell is tripled by the CDW order so this state can exist without extra topological order.

{\bf -- Translation breaking enforced by filling:} Importantly, when the translation symmetry of the moir\'{e} superlattice is preserved, it is not possible  (at least within CF mean-field theory) to have a direct transition from a state that fills a CF band with Chern number $C_{cf} = 2$ and correspondingly Hall conductivity $\sigma_H=-2/3$, to states with $C_{cf}=1,0$ and $\sigma_H=-1, 0$. Since the effective field seen by the CFs is $\phi_{cf} = 1/3$ flux quanta through each moir\'{e} plaquette, the CFs obey the magnetic translation symmetry: $T_1T_2=T_1^{-1}T_2^{-1}e^{\ii 2\pi/3}$ for the two elementary translations $T_{1,2}$ that enclose a moir\'{e} unit cell. The mean-field spectrum of the CFs will be $3$ fold degenerate as the magnetic translation admits representations with minimal dimensions of $3$.~\cite{burkov2005superfluid} 

Therefore the translation symmetry guarantees that the change of the Chern number $\Delta C_{cf}$ across a transition must have $\Delta C_{cf}$ equal to multiples of $3$, as it is given by the number of gapless Dirac cones in the spectrum. Hence if the transitions $\sigma_H=-2/3\rightarrow -1$ and $\sigma_H=-2/3\rightarrow 0$ are described by changing the band Chern number of the composite fermions, they can only occur when the translation symmetry of the moire lattice is broken, i.e. they can only occur when there is a background charge density wave order. In particular, the simplest scenario of CDW is to triple the unit cell, rendering the magnetic translation trivial and permitting a direct transition with $\Delta C_{cf}=-1,-2$ etc.  
%In general a transition with gap closing and reopening will lead to a change of Chern number by $1$, without any symmetries. 
In the next section, we shall demonstrate explicitly direct transition from $\sigma_H=-2/3$ FQAH-crystal to either $\sigma_H=0$ CDW or $\sigma_H=-1$ QAH-crytsal.    
%will see that a direct transition $\sigma_H=-2/3\rightarrow 0$ requires certain extra discrete symmetry to guarantee a two-fold degeneracy of gapless Dirac cones in the folded BZ due to the background CDW. %For example, the ideal Kane-Mele model of the homobilayer twisted TMD system has a combination of reflection that exchanges the two valleys,
%(that exchanges the sublattice of the honeycomb lattice) 
%and the time reversal, i.e. $R \otimes T$ as a good symmetry\cite{wu2019topological,reddy_2023_fqah}. Hence a Dirac cone at the $K$ point in the folded Brillouin zone will have a $R \otimes T$ counterpart at $K'$.

%{\bf --- State 3, Integer QAH state with $\sigma_H = 1:$ }

%The parton picture: the bosonic parton forms $\nu = -1/2$ state; the fermion partons fill two bands with Chern numbers $+1, 0$. 

%Alternatively this can be viewed as holes at filling $-1$, forming a $\nu=-1$ integer quantum hall(IQH) state, together with electrons at filling $1/3$, forming a trivial insulating state.

%Composite fermion picture: 

%In addition to tuning the states of $f$, we could alternatively tune the states of $\Phi$
%\end{enumerate}

\begin{widetext}
\begin{center}
\begin{table}
    \begin{tabular}{|c|c|c|c|c|}
    \hline
 Phase   & CF & Parton $f$ ($\Phi$ in Laughlin $-\frac{1}{2}$ state) & Electron+holes($\nu=-1$ IQH) & K matrix\\ \hline
$\sigma_H=-\frac{2}{3}$ & $C_{cf}=2$ & $C_f=2$ & Electron in $\frac{1}{3}$ Laughlin& $\begin{pmatrix} -2&0&0&1\\ 0&1&0&-1\\0&0&1&-1\\1&-1&-1&0\end{pmatrix}\simeq \begin{pmatrix} -1&0&0&0\\ 0&3&0&0\\0&0&0&-1\\0&0&-1&0\end{pmatrix}$\\ \hline
$\sigma_H=0$ & $C_{cf}=0$ & $C_f=0$ & Electron in $\nu=1$ IQH& $\begin{pmatrix} -2&0&0&1\\ 0&1&0&-1\\0&0&-1&-1\\1&-1&-1&0\end{pmatrix}\simeq \begin{pmatrix} -1&0&0&0\\ 0&1&0&0\\0&0&0&-1\\0&0&-1&0\end{pmatrix}$\\ \hline
$\sigma_H=-1$ & $C_{cf}=1$ & $C_f=1$ & Electron in trivial insulator & $\begin{pmatrix} -2&0&1\\ 0&1&-1\\1&-1&0\end{pmatrix}\simeq \begin{pmatrix} -1&0&0\\ 0&0&1\\0&1&0\end{pmatrix}$\\ \hline %\label{Kmatrix}
\end{tabular}
\caption{Summary of the $3$ formalisms that describes the three states with $\sigma_H=-2/3,-1,0$ respectively. When the composite fermions (CF) fill Chern bands with total Chern number $C_{cf}$, the physical Hall conductivity is $\sigma_H=\frac{C_{cf}}{1-2C_{cf}}$.\label{tab:summary}}
\end{table}
\end{center}
\end{widetext}

\subsection{Phases tuned by bosonic parton $\Phi$}

\begin{figure}[tb]
\begin{center}
\includegraphics[width=0.85\linewidth]{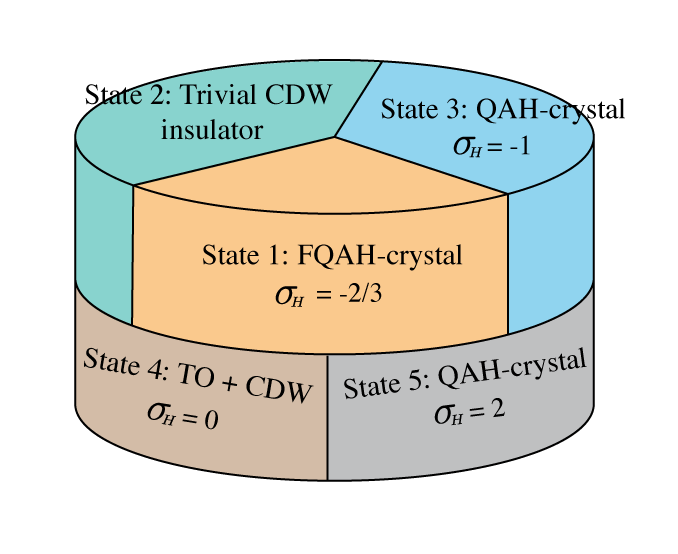}
%\vspace{-.1in}
\end{center}
\caption{A schematic global phase diagram in terms of the parton construction, tuned by both physics of bosonic parton $\Phi$ (vertical direction) and fermionic parton $f$ (horizontal directions) starting from the FQAH-crystal. We discuss interesting critical theories among the phases shown in Sec. \ref{sec:transition}.}
\label{phasedia}
\end{figure}

In all the three states constructed in the last subsection, the bosonic parton $\Phi$ always form a $\nu = - 1/2$ bosonic Laughlin state. Starting with the FQAH-crystal state with $\sigma_H = - 2/3$, two more states can be constructed by changing the physics of the bosonic parton $\Phi$. These states/phases are summarized in the global phase diagram Fig.~\ref{phasedia}. 

%In state 1 the bosonic parton $\Phi$ is in a $\nu = - 1/2$ bosonic fractional QH state, and the fermionic parton $f$ will fill two bands with Chern number $+1$, this is the FQAH state with total Hall conductivity $\sigma^H = - 2/3$. 

%Phase 2 is the trivial insulator phase discussed in the previous section with a density wave that spontaneously breaks the translation symmetry. Starting with phase 1, phase 2 can be constructed by making the fermionic partons filling two bands with Chern numbers $\pm 1$ respectively, while keeping the state of $\Phi$ unchanged. 

One such state is an insulator without any Hall conductivity, but it has a neutral topological order and neutral chiral edge states, leading to quantized thermal Hall effect. This quantum thermal Hall insulator can be obtained from 
the FQAH-crystal, by driving the bosonic parton $\Phi$ into a trivial insulator. When $\Phi$ is in a trivial insulator, there is no nontrivial response to the external electromagnetic field as $\Phi$ is the parton that carries the electric charge. However, this state must still have a nontrivial topological order, as the $K$ matrix of this state corresponds to Eq.~\ref{Kmatrix1} after removing the components that involve gauge field $b$. The determinant of the remaining $3 \times 3$ $K$ matrix is $2$, and it is equivalent to a simple semion topological order. The semion topological order can also be revealed by integrating out $a_1$ and $a_2$, which yields a level-2 CS term for the gauge field $a$. This semion topological order with zero Hall conductivity is one of the states discussed in Ref.~\onlinecite{song2023phase}.

The other state is an QAH-crystal state with Hall conductivity $\sigma_H = + 2$. This state can be constructed from FQAH-crystal by driving the bosonic parton $\Phi$ into a ``superfluid" state. In the condensate of $\Phi$, the hole operator $c$ is identified with $f$, and since $f$ fills two bands with Chern number $+1$, this leads to a QAH-crystal state with Hall conductivity $\sigma_H = 2$. An QAH state without topological order is possible as we assumed a background CDW that triples the unit cell. 

%\begin{figure}[tb]
%\begin{center}
%\includegraphics[width=0.75\linewidth]{phasedia.png}
%\vspace{-.1in}
%\end{center}
%\caption{A global phase diagram in terms of the parton construction, tuned by both physics of bosonic parton $\Phi$ (polar angle direction) and fermionic parton $f$ (radial direction of transition between state 2 and state 1 or 5). We discuss interesting critical theories among the phases shown in sec \ref{sec:transition}.}
%\label{phasedia}
%\end{figure}

\section{Quantum Phase Transitions}

\label{sec:transition}
So far, we have constructed five different states centered around the ``2/3" state, in a phase diagram tuned by mean field physics of the bosonic parton $\Phi$ and the fermionic parton $f$. These states can also be equally well constructed through other formalisms, including the composite fermions and flux attachment. In this section, we discuss the quantum phase transitions between these states. Here we stress that the FQAH state we start with is in fact a FQAH-crystal, while the starting point of Ref.~\onlinecite{song2023phase} was a FQAH state without spontaneous translation symmetry breaking. Hence different proximate phases and phase transitions are obtained in these two papers. For example, in our current case the most natural $\sigma_H = 0$ insulator state next to the $\sigma_H = -2/3$ state is a trivial insulator with CDW, while in Ref.~\onlinecite{song2023phase} the $\sigma_H = 0$ state has a topological order that would lead to nontrivial thermal Hall signal. 

%{\color{blue} The parton construction for and transitions out of $2/3$ state were discussed in ref \onlinecite{song2023phase}, where mainly the states of $\Phi$ are tuned to describe FQAH and proximate phases. Here besides the ingredient of entertwined CDW order with FQAH, we offer a complementary picture and focus on tuning the $f$ phase to control the phase diagram. The critical theories constructed from tuning $f$ or $\Phi$ are formally related by boson-fermion duality and we comment on physically interesting examples below where appropriate.}

%\begin{enumerate}
\subsection{$\sigma_H = -2/3 \rightarrow  0$ transition}

To drive a transition between the FQAH-crystal state with $\sigma_H = -2/3$ and a trivial CDW insulator with $\sigma_H = 0$, we can keep the bosonic parton $\Phi$ in the $\nu = - 1/2$ state unchanged, and only change the total Chern number of the fermionic parton bands from $C = 2$ to $C = 0$, which can be realized by changing one of the occupied bands from $C= +1$ to $C = -1$. If there is a direct transition between these two states, it must involve two Dirac fermions at low energy. 
%and these two Dirac fermions are coupled with a level-$1/2$ Chern-Simons(CS) term accounting for the remaining filled band with $C=1$.
The complete critical theory reads:
\beqn
    \mathcal L_{1-2} &=& \sum_{i=1,2}\bar \psi_i \gamma \cdot (\ii\partial - a)\psi_i + m\bar\psi\psi + \frac{1}{2\pi}ad(\alpha - b) \cr
    &+& \frac{1}{4\pi}\alpha d\alpha -\frac{2}{4\pi}bdb + \frac{1}{2\pi}Adb, \label{L12}
\eeqn
where $\alpha,b$ are the dual fields associated with the filled $C=1$ band of $f$ (hence the $U(1)_1$ for $\alpha$) %\cx{(CX: I would prefer to call $\alpha$ and $b$ as the dual of the $C=1$ $f$, and $\Phi$, rather than auxiliary gauge field)} 
and the boson $\Phi$ currents, respectively. 
%Hence $db$ couples to $A-a$, while $d\alpha$ couples to $a$, given that $f,\Phi$ couple to $a,A-a$, respectively. 
Two Dirac fermions must exist at low energy at the transition for one $f$-band to change from $C=1$ ($m>0$) to $C=-1$ ($m<0$).

The theory Eq.~\ref{L12} can be simplified at the cost of losing the proper quantization of the CS terms. Integrating out $\alpha$ generates $- 1/(4\pi)ada$.%~\footnote{We keep in mind that $a$ couples to a fermion, hence it is a spin$_C$ gauge field.}, 
and integrating out $b$ generates $1/(8\pi)(A - a)d(A - a)$. The simplified theory then reads:
\beqn
    \mathcal L_{2;-\frac{1}{2}} &=& \sum_{i=1,2}\bar \psi_i \gamma\cdot (\ii\partial - a)\psi_i + m\bar\psi\psi - \frac{1}{8\pi}ada\nonumber \cr
    &-& \frac{1}{4\pi}adA + \frac{1}{8\pi}AdA.
    \label{eq1-2}
\eeqn
We use the notation $\mathcal L_{N_f;k}$ to label the QED Lagrangian with $N_f$ flavors of Dirac fermions and a Chern-Simons term at level $k$. 

The degeneracy of the two Dirac fermions can be guaranteed by extra discrete space-time symmetries. In the absence of displacement field, the entire homobilayer twisted TMD moir\'{e} system has a $C_{2y}$ symmetry, a two-fold rotation along the vertical axis in Fig.~\ref{fig:cdw_bz}, as well as a time-reversal symmetry $\mathcal{T}$. Both symmetries exchange the two valleys. Hence, each valley of the system holds a composite symmetry of $C_{2y} \mathcal{T}$. This composite symmetry sends $(k_x,k_y)\rightarrow (k_x,-k_y)$.
%and enforces $2$ degenerate Dirac cones. 
The degeneracy of the two Dirac fermions that is needed for a direct ``$\sigma_H = - 2/3 \ra \sigma_H = 0$" transition in our set-up depends on the type of CDW order. For example, if the CDW is a stripe order along the $y$ direction with modulation along the $x$ direction as shown in Fig.~\ref{fig:stripe_bz}, the two Dirac points could still be located at $(k_x,-k_y)$ and $(k_x,-k_y)$ points of the BZ, and their degeneracy is still protected by the $C_{2y}\mathcal{T} $ symmetry. In contrast, if there is a $\sqrt{3} \times \sqrt{3}$ CDW order with $C_3$ symmetry shown in fig \ref{fig:cdw_bz}, there is no scenario where $C_3, C_{2y} \mathcal{T}$ together protect two and only two degenerate Dirac cones. Here we note that an out-of-plane displacement field in principle breaks the $C_{2y}$ symmetry as it exchanges the two layers, hence under a displacement field the transition may split into two. 

Although the microscopic symmetry $C_{2y}$ of the system may be broken by a displacement field, extra effective symmetries may still exist in the physics of the moire minibands. For example, there is an extra discrete symmetry of the Hamiltonian that describes one valley of the system~\cite{wu2019topological}, which is a composite of $R_x: y \rightarrow -y$, and a ``time-reversal" that acts on this one-valley Hamiltonian. This symmetry still exists with the presence of the displacement field. If this symmetry is a good approximate symmetry of the moire miniband, it can also protect the degeneracy of two Dirac points and a direct transition of changing Chern number by $2$, as was observed in model studies in Ref.~\onlinecite{wu2019topological,reddy2023fractional}.
%The direct transition of ``$\sigma_H=-2/3\rightarrow \sigma_H = -1$", to be discussed in the following section, does not need symmetry protection as it involves a single Dirac cone.
%To explicitly see that the theory indeed describes the transition between $\sigma_H=-2/3,0$ phases, for $m>0,m<0$, respectively, we could integrate out the $\psi$'s and introduce an auxiliary field $\beta$: this leads to $-sgn(m)/(4\pi)\beta d\beta+1/(2\pi)ad\beta$. Further integrates out $\beta$, one has for the Lagrangian 
%\begin{align}
%    \mathcal L_m=\frac{sgn(m)+1/2}{4\pi}ada-\frac{1}{4\pi}adA-\frac{1}{8\pi}AdA.
%\end{align}

%For $m>0$, we have CS term for $a$ at level $-3/2$,%this is represented by the K matrix $K=-\begin{pmatrix} 2&1\\ 1&2\end{pmatrix}$ with a charge vector $l=(0,1)^T$. 
%hence the hall conductivity $\sigma_H=-2/3$. For $m<0$, integrating out $a$ gives trivial response to $A$. %the CS term for $a$ cancels out and integrating out $a$ sets $b=0$. 
%The action describes a trivial insulator with Maxwell action of electromagnetic field $A$.

An alternative description for the same transition from the $\sigma_{H}=-2/3$ state to the $\sigma_H=0$ state can be constructed using the ``electron-hole picture": The $\sigma_H = - 2/3$ state can be viewed as a composition of holes at the $\nu=-1$ IQH state, and electrons in the Laughlin $\nu = 1/3$ state. To drive a transition to the $\sigma_H = 0$ state we need a transition of the electrons from the $\nu = 1/3$ state to the $\nu=1$ IQH state. 
%The entire system then has the correct Hall response for the two phases. 
%The transition is given by the electrons at filling $1/3$, transitioning from Laughlin $1/3$ to $\nu=1$ IQH.
The critical theory reads:
\beqn
    \mathcal L_{1-2;eh} &=& \sum_{i=1,2}\bar \psi \gamma \cdot (\ii\partial - a)\psi + m\bar\psi\psi-\frac{1}{2\pi}ad\beta \cr &+& \frac{2}{4\pi}\beta d \beta+\frac{1}{2\pi}Ad\beta + \frac{1}{4\pi}AdA, \label{L12eh}
\eeqn 
%where the Dirac fermions appear to describe the Chern number of fermionic parton $f$ to change from $C=1, (m>0)$ to $C=-1,(m<0)$. The field $b$ again describes the $\Phi$ Laughlin $1/2$ state. 
The last term $+ \frac{1}{4\pi}AdA$ accounts for the hole $\nu = -1$ state that stays unchanged throughout the transition. The rest of the Lagrangian describes the transition of the electrons from the $\nu =1/3$ state to a $\nu = 1$ IQH state. When the $2$ Dirac cones are gapped out by a mass, integrating out the $\psi$'s give
\beqn
        \mathcal L_m &=& \frac{- \sgn(m)}{4\pi}ada-\frac{1}{2\pi}ad\beta +\frac{2}{4\pi}\beta d\beta \cr
        &+& \frac{1}{2\pi}Ad\beta + \frac{1}{4\pi}AdA.
\eeqn
It is straightforward to verify that for $m > 0$ ($m < 0$), the theory describes states with Hall responses $\sigma_H=-2/3$ ($\sigma_H = 0$) respectively. Starting with Eq.~\ref{L12eh}, the theory simplifies again to $\mathcal L_{2;-\frac{1}{2}}$ after integrating out $\beta$.

%Similarly, %{\color{blue} {It it unclear whether the $2$ theories $\mathcal L_{1-2},\mathcal L_{1-2;eh}$ are dual to each other, as they seem to describe the same continuous transition whose proximate phases are identically state 1,2.}}

\subsection{$\sigma_H = -2/3\rightarrow  -1$ transition}

From the parton picture, this transition involves changing a fermion $f$'s state from integer quantum Hall state with $\nu = +1$ to $\nu = 0$. This transition can be described by QED with one Dirac fermion in the infrared, the critical theory reads:
\beqn
    \mathcal L_{1-3} &=& \bar \psi \gamma\cdot (\ii \partial -a)\psi+m\bar\psi\psi - \frac{1}{8\pi}ada+\frac{1}{2\pi}ad(\alpha-b) \cr
    &+& \frac{1}{4\pi}\alpha d\alpha - \frac{2}{4\pi}bdb+\frac{1}{2\pi}Adb, \label{eq13a}
\eeqn
%where the single Dirac fermion is involved to describe Chern number change $C=1, (m>0)$ to $C=0,(m<0)$. 
We have added $- \frac{1}{8\pi}ada$ to properly regularize a single Dirac cone, which arises from another massive Dirac fermion which must exist in the same band as $\psi$. Here, the single Dirac fermion is written in the convention that, by changing the sign of $m$, $\psi$ would generate a level $\pm 1/2$ CS term for $a$. Integrating out $b,\alpha$, we obtain a simplified theory (again at the cost of not properly quantizing the CS term)
\beqn
    \mathcal L_{1; -1} &=& \bar \psi \gamma \cdot (\ii \partial - a)\psi + m\bar\psi\psi - \frac{1}{4\pi}ada\nonumber \cr
    &+& \frac{1}{8\pi}AdA - \frac{1}{4\pi}adA.
    \label{eq1-3}
\eeqn

Similarly, when the  Dirac cone is gapped out by a {\it positive} mass term $m \bar{\psi}\psi$, integrating out the $\psi$'s give
\begin{align}
        \mathcal L_{m>0}= - \frac{3}{8\pi}ada + \frac{1}{8\pi}AdA-\frac{1}{4\pi}adA,
\end{align}
which generates Hall conductivity $\sigma_H=-2/3$. While $m<0$ one has
\begin{align}
        \mathcal L_{m<0}= - \frac{1}{8\pi}ada + \frac{1}{8\pi}AdA-\frac{1}{4\pi}adA,
\end{align}
and integrates out $a$ leaves $\frac{1}{4\pi}AdA$, describing
a state with $\sigma_H = -1$.

From standard boson-fermion duality~\cite{wudual,SEIBERG2016} %\cx{(I appreciate that our paper is cited here, but I think a better reference would be the paper by Senthil, Nati, et.al.)}
, the critical theory $\mathcal L_{1;-1}$ is dual to 
\beqn
    \mathcal L_{1;-1} &\leftrightarrow& |(\partial - \ii \beta)\phi|^2 + \frac{1}{4\pi}\beta d\beta + \frac{1}{2\pi}ad\beta - \frac{1}{8\pi}ada\nonumber \cr
    &+& \frac{1}{8\pi}AdA - \frac{1}{4\pi}adA. 
\eeqn
Here $\beta$ is another gauge field that couples to the dual bosonic field $\phi$. One could verify that the massive and condensed phase of $\phi$ corresponds to $\mathcal L_{m>(<)0}$ respectively, yielding $\sigma_H=-2/3,-1$. One can also directly perform the duality transformation from Eq.~\ref{eq13a}.

Integrating $a$ in the dual bosonic theory leaves
\begin{align}
    \mathcal L_{1;-1}\leftrightarrow |(\partial - \ii \beta)\phi|^2 + \frac{3}{4\pi}\beta d \beta - \frac{1}{2\pi}\beta d A + \frac{1}{4\pi}AdA. \label{dual}
    \end{align}
This Chern-Simons-matter theory with $\phi$ coupled to a $\U(1)$ gauge field with a CS term with level-3 is the standard theory that describes a transition between a trivial insulator and a fractional quantum Hall state with three-fold topological degeneracy~\cite{wenwu}. Combined with the last term $\frac{1}{4\pi}AdA$ which corresponds to an extra $\nu=-1$ IQH layer, the theory describes a transition between states with $\sigma_H = - 2/3$ and $\sigma_H = -1$.
This FQAH-crystal to QAH-crystal transition also admits another description in terms of bosonic partons, which is a modified version of the FQAH to QAH+CDW transition discussed in Ref.\onlinecite{song2023phase} driven by the condensation of $3$ `vortex' fields coupled to a $\U(1)_3$ Chern-Simons term. It is worth noting that this modified vortex condensation theory takes the same form as Eq.~\ref{dual}. \footnote{The multiple vortex fields are required as the $2/3$ state there in Ref.\onlinecite{song2023phase} preserved translation, hence vortex degeneracy enforced by fractional filling. When CDW orders are assumed here and integer filling is reached for enlarged unit cell, the transition is modified to a single vortex $\phi$ condensation transition discussed here.}

%gives a natural physical interpretation of composition of $\nu=-1$ IQH layer from the last term $\frac{1}{4\pi}AdA$ and a transition between $\nu=1/3$ Laughlin state (given by the $\frac{3}{4\pi}bdb$) and a trivial insulator. This again confirms the electron-hole construction we present.

\subsection{$\sigma_H = -2/3 \rightarrow + 2$ transition}

Another potentially direct transition is between state 1 and 5, i.e. a transition from a FQAH-crystal state with $\sigma_H = - 2/3$ to an QAH-crystal state with $\sigma_H = + 2$. In the parton construction, this requires changing the state of $\Phi$ from a $\nu = - 1/2$ Laughlin state to a ``superfluid" state. This transition of $\Phi$ was discussed in Ref.~\onlinecite{barkeshli2014continuous,barkeshli2012continuous}, and it is described by a QED with two flavors of Dirac fermions and a CS term at level $-1$. In our notation, the critical theory of $\Phi$ is described by a Lagrangian $\cL_{2; -1}$, and the Dirac fermions are charges of the gauge field $b$, i.e. the dual of the current of $\Phi$. To describe the transition between states 1 and 5, we need to couple $b$ to several other gauge fields $a$, $a_i$ as in Eq.~\eqref{cs1}. After integrating out the $a$ and $a_i$, we arrive at the critical theory between state 1 and 5: \beqn && \cL_{1-5} =\tilde\cL_{2; - 1/2} = \cr\cr && \sum_{i = 1,2} \bar{\chi}_i \gamma \cdot (\ii \partial - b) \chi_i + m \bar{\chi}\chi - \frac{1}{8\pi} bdb + \frac{1}{2\pi} A db. \eeqn We note that here the Dirac fermion $\chi_i$ is different from the fermions in the previous sections, as it is charged under $b$(hence the critical theory $\tilde\cL_{2,-1/2}$ is different from $\mathcal L _{2;-1/2}$ in Eq. \eqref{eq1-2} with different charge assignment). The degeneracy of two Dirac cones is again protected by $C_{2y} \mathcal{T}$ in a stripe order.

Integrating out the fermion $\chi_i$, we obtain the following action \beqn \cL_m = - \frac{\sgn(m)}{4\pi} b db - \frac{1}{8\pi} bdb + \frac{1}{2\pi} A db. \eeqn It is straightforward to verify that, for $m > 0$ ($m < 0$), the final Hall conductivity is $\sigma_H = - 2/3$ ($\sigma_H = +2$). 

\subsection{Transitions involving quantum thermal Hall insulator}

Ref.~\onlinecite{song2023phase} proposed a proximate insulating phase of the $\sigma_H=-2/3$ FQAH state to be one with vanishing Hall response and a neutral topological order (TO) described by the $\U(1)_2$ CS terms. In the current framework, this state could be obtained by putting the bosonic parton $\Phi$ in a Mott insulator state. Formally, this amounts to eliminating the dual $b$ of the boson current, i.e. setting $b=0$, in the construction of the $2/3$ state in Eq.~\eqref{cs1}. Physically, it means that the bosonic sectors are trivially gapped in the low energy. Integrating out $a$ then sets $a_1=a_2$, and one gets a $\U(1)_2$ CS coupling of an internal gauge field, describing the neutral TO, signified by a quantized thermal hall response nevertheless.

The transition from quantum thermal Hall insulator to $\sigma_H=-2/3$ FQAH-crystal or $\sigma_H=+2$ QAH-crystal, can hence be obtained by tuning the $\Phi$ out of the Mott phase to a Laughlin $-1/2$ state (for state 1), or a superfluid (for state 5), respectively. 

The Mott to Laughlin $-1/2$ transition is realized by condensing the ``vortices" of the bosonic partons% \cx{(I think we had better use ``bosonic parton" rather than ``composite boson", as we have not used the notion of composite bosons in the entire paper)} 
in the Laughlin $-1/2$ state\cite{song2023phase}.
The critical theory describing the transition from state 1 to state 4 hence reads 
\begin{align}
    \mathcal L_{1-4}=|(\partial - \ii b)\Phi_v|^2 - \frac{2}{4\pi}bdb + \frac{1}{2\pi}Adb\nonumber\\
    \frac{1}{2\pi}ad(b-a_1-a_2)+\sum_{i=1,2}\frac{1}{4\pi}a_ida_i.
\end{align}
The first term describes the condensation of the vortices $\Phi_v$, which is indicated by its coupling to $b$, whose flux equals the boson density. As $\Phi_v$ condenses, the field $b$ acquires a gap and could be ignored. Hence, one arrives at the quantum thermal Hall insulator with $\sigma_H=0$. An insulator phase of $\Phi_v$ just leaves the Lagrangian for the $2/3$ state Eq.~\eqref{cs1}. A similar transition was discussed in ref \onlinecite{song2023phase}, albeit with $3$ vortex fields enforced by fractional filling $2/3$. %We discussed a simpler theory with one vortex field due to the presence of CDW order postulated in the FQAH state.}

The transition from quantum thermal Hall insulator to QAH-crystal with $\sigma_H=+2$ is given by the standard boson Mott-superfluid transition,
\begin{align}
    \mathcal L_{4-5}=|(\partial - \ii a -\ii A)\Phi|^2-
    \frac{1}{2\pi}ad(a_1+a_2)+\sum_{i=1,2}\frac{1}{4\pi}a_ida_i.
\end{align}
When $\Phi$ is gapped, the remaining last $2$ terms describes the neutral topological order. The condensation of $\Phi$ sets $a=-A$, and the CS terms of $a_i$'s then give a Hall conductivity $\sigma_H=2$.

\section{Summary}

In this work we discussed the phase diagram centered around a FQAH-crystal state with $\sigma_H = -2/3$ state at filling $-2/3$, motivated by recent experiments.%, and numerical work that observed a coexistence between FQAH state and CDW (\cx{reference?}).  
 Various phases and phase transitions can be obtained by tuning the physics of the bosonic and fermionic partons, including a direct transition between the $\sigma_H = - 2/3$ state and a trivial insulating state with $\sigma_H = 0$ observed in recent experiments. Interestingly, we also find a direct transition between the $\sigma_H = - 2/3$ FQAH state and a $\sigma_H=-1$ QAH state. 

Our formalism and conclusions can easily be generalized to other FQAH states. For example, if the bosonic parton $\Phi$ forms a $\nu = - 1/p$ Laughlin state (with even integer $p$), and the fermionic parton (or composite fermion) $f$ fills Chern bands with total Chern number $C_{cf}$, we would end up with an FQAH state with Hall conductivity \beqn \sigma_H = \frac{C_{cf}}{1 - p C_{cf} }. \eeqn 

In particular, when $p=-2,C_{cf}=1$, i.e. the composite fermions fill a Chern band with $C_{cf}=1$, one constructs a Laughlin state with $\sigma_H=1/3$. When the CFs go through a transition from $C_{cf}=1\rightarrow 0$, the electronic state transitions from $\sigma_H=1/3\rightarrow 0$. The critical theory is similar to that of $\sigma_H=-2/3\rightarrow -1$ with the Lagrangian $\mathcal L_{1;-1}$ in Eq.~\eqref{eq1-3}, with an only difference of an integer quantum hall layer described by $-1/(4\pi)AdA$. The critical theory for the transition hence reads
\begin{align}
   \mathcal L_{1/3\rightarrow 0}=\mathcal L_{1;-1}-\frac{1}{4\pi}AdA.
\end{align}

More states and phase transitions can be constructed by changing the ``mean field states" of $\Phi$ and $f$. This general construction will be useful to understand the growing number of FQAH states~\cite{lu2023fractional} observed in this rapidly developing field. 

{\it Acknowledgement} We thank insightful discussions with Xiaodong Xu. LF thanks related collaborations with Aidan Reddy, Hart Goldman, Nisarga Paul, Ahmed Abouelkomsan and Emil Bergholtz.  
XYS thanks collaborations and discussions with T. Senthil and Y-H Zhang. XYS is supported by the Gordon and Betty Moore Foundation EPiQS Initiative through Grant No.~GBMF8684 at the Massachusetts Institute of Technology. CMJ is supported by a faculty startup grant at Cornell University. LF is supported by the Air Force Office of Scientific Research (AFOSR) under Award No. FA9550-22-1-0432.  
CX is supported by the Simons foundation through the Simons Investigator program.


%\bibliography{main}

\end{document}